\documentclass{elsart3p}
\usepackage{graphicx}
\usepackage{color}

\usepackage{amssymb}

\begin{document}
\newcommand{\BSCCO}{{Bi$_2$Sr$_2$CaCu$_2$O$_{8+x}$ }}
\begin{frontmatter}



\title{Signatures of modulated pair interaction in cuprate superconductors}


\author[a]{T.S. Nunner,}
\author[b]{P.J. Hirschfeld,}
\author[b]{B.M. Andersen,}
\author[b]{A. Melikyan,}
\author[c]{and K. McElroy}

\address[a]{Institut f\"ur
Theoretische Physik, Freie Universit\"at Berlin,
 Arnimallee 14, 14195 Berlin, Germany}
 \address[b]{Department of Physics, University of Florida, Gainesville, Florida 32611-8440, USA}
 \address[c]{Material Sciences Division, Lawrence Berkeley National Laboratory, Berkeley, California 94720, USA }

\begin{abstract}
Recent low-temperature scanning tunnelling spectroscopy
experiments on the surface of \BSCCO have revealed a strong positive
correlation between the position of localized resonances at -960 meV
identified with interstitial oxygen dopants and the size of the
local spectral gap. We review efforts to understand these
correlations within a model where the dopants
modulate the pair interaction on an atomic scale. We
provide further evidence for this model
by comparing the correlations between the dopants and the
local density of states with experimental results.
\end{abstract}

\end{frontmatter}

\section{Introduction}
\label{introduction} High-resolution scanning tunnelling (STS)
spectroscopy on BSCCO-2212 has provided some of the most provocative data on
cuprate superconductors in recent years, including local
information on impurity states, charge ordering, quasiparticles in
the vortex state, and nanoscale inhomogeneity.
In particular the observation of nanoscale inhomogeneity in the
electronic structure of BSCCO-2212 has raised great
interest due to its possible relevance to understanding  how the
cuprates evolve from the doped Mott insulator to the
superconducting state.
 Recently, STS provided intriguing results
which may bear on this question, by the observation of random
localized resonances at a bias of -960 meV which were argued
to be states associated with interstitial oxygen
atoms\cite{DavisScience05}. These defects  dope the  BSCCO-2212
crystals, along with a rather complex mix of other
dopants\cite{Eisaki}.
From these experiments a strong and positive correlation between
the magnitude of the gap and the O dopant positions has been
inferred by McElroy {\it et al.}~\cite{DavisScience05}. This result
was unanticipated, since it had been suggested~\cite{Martin} that each O might
overdope the crystal locally by providing two holes, corresponding
to a smaller local gap in analogy with the average gap decrease
with overdoping known from tunnelling experiments~\cite{Miyakawa}. Additional
salient characteristics of the STS-spectra are their remarkable
particle-hole symmetry and their homogeneity at low energies.

Nunner {\it et al.}~\cite{TSNunner:2005} subsequently argued that
these main characteristics of the STS-spectra cannot be explained
by a model where the dopant atoms act simply as potential
scatterers. Instead they suggested that the hypothesis of enhanced
pairing around each O-dopant can explain the observed
correlations, in particular the negative correlation 
of the coherence peak height with the magnitude of the gap, i.e. the
fact that the coherence peaks are suppressed in regions of large
gap and enhanced in regions of small gap. The underlying physics
within this model is Andreev-scattering of the quasiparticles at
order parameter modulations. In regions of suppressed order
parameter a new resonance forms below the gap
edge~\cite{TSNunner:2005} which has been used to explain the
extremely sharp coherence peaks observed by Fang {\it et
  al.}~\cite{Fang} in small gap regions. In regions of
enhanced order parameter exactly the opposite happens and the coherence peaks are
strongly suppressed~\cite{TSNunner:2005}. The fact that the scattering
mechanism is of Andreev type can also account for the
remarkable particle-hole symmetry of the STS-spectra and for the observed homogeneity
at low energies since nodal quasiparticles are less affected by
order parameter variations than antinodal quasiparticles. In addition,
modulation of the charge is found to be quite weak,
in agreement with experiment~\cite{DavisScience05}. None of these
effects can be found in models of conventional potential disorder.

Further comparison of the pair disorder model with experiment
revealed a natural explanation of the relative  strengths of  the
quasiparticle interference peaks observed in Fourier transform
scanning tunnelling  spectroscopy (FT-STS)~\cite{TSNunner:2005b}
and showed that the broadening of thermodynamic transitions
induced by the gap modulations observed by STS was compatible with
specific heat measurements on BSCCO-2212~\cite{BMAndersen:2006}.
This serves as an additional evidence that the nanoscale inhomogeneity
observed by STS is indeed characteristic of the bulk in the BSCCO-2212
system.

The success of this hypothesis raises the question of whether one
can learn about the origins of pairing by studying what modulates
it directly~\cite{ZWang:2006}.  If one were to know the changes in
electronic structure induced by a dopant or other impurity, and
could establish a direct link to the change in the superconducting
order parameter, one would have  valuable
information with which to constrain theories of superconductivity.
A possible microscopic origin of the modulated pair interaction might
be the distortion of the crystal lattice around each O interstitial.
These local lattice distortions might change local electron-phonon
interaction matrix elements or modulate the nearest neighbor
hopping $t$ and as a consequence the superexchange constant  $J$
between nearest neighbor Cu-atoms (in a one-band model e.g.
$J=4t^2/U$). Within a strong-coupling picture
one might assume that the pair interaction is provided by the
superexchange $J$ and any modulations of $J$ would
directly translate into a modulated pair interaction. This scenario has
been further worked out by Zhu~\cite{JXZhu:2005} with results 
similar to the ones found by Nunner {\it et al.}~\cite{TSNunner:2005}.

To understand the detailed changes in electronic structure around
a dopant atom, one can perform an ab initio density functional
theory (DFT) calculation of the structure with and without the
interstitial present.  A first attempt of this type for an oxygen
interstitial in the BSCCO-2212 system was made by He
{\it et al.}~\cite{YHe:2006}, who found that the lowest energy location for
such a dopant was between the BiO and SrO planes.  Remarkably
enough, the DFT shows the presence of a narrow band of states near
-1~eV induced by the dopant oxygen, as observed in
experiment~\cite{DavisScience05}.  These states have
O(dopant)2$p_z$ character, possibly explaining their strong
coupling to the STM tip states, but mix with Bi and O(Bi) states
as well.  With the results of these calculations, one may now
attempt to correlate changes in gap observed by STS with changes
in fundamental electronic structure parameters like $t,$ $t'$,
electron-phonon coupling constants and Madelung energies.

In the remainder of the paper we provide further evidence in favor
of the hypothesis that the oxygen dopant atoms modulate the pair
interaction locally rather than to act solely as potential
scatterers. For this purpose we calculate the correlation between
the local density of state (LDOS) and the modulated pair
interaction/disorder potential and compare it with experimental
results for the correlation of the LDOS with the positions of the
oxygen dopant atoms as found by McElroy {\it et
al.}~\cite{DavisScience05}.

\section{Models of inhomogeneity}

Our calculations are based on the standard mean field Hamiltonian for
a singlet $d$-wave superconductor
\begin{eqnarray}
\label{eq:hamiltonian} 
\!\!\hat{H}\!\!=\!\!\!\sum_{ij \sigma}\! 
\left( t_{ij} \!\!+\!\! \delta_{ij}(V_i\!\!-\!\!\mu) \right)
\hat{c}_{i\sigma}^\dagger
\hat{c}_{j\sigma} \!+\!\! \sum_{\langle ij
\rangle} \!\! \left( \!\Delta_{ij} \hat{c}_{i\uparrow}^\dagger
\hat{c}_{j\downarrow}^\dagger \!\!+\! \mbox{h.c.} \! \right)\!\!,\!
\end{eqnarray}
where we keep only nearest $t$ and next-nearest $t^\prime=-0.3t$
neighbor hopping, with  $\mu=-t$  to model the Fermi surface of
BSCCO near optimal doping. In the following we consider two models of
how the oxygen dopant atoms create inhomogeneity in the local electronic
structure. In model (i) we follow the conventional assumption that
each oxygen dopant acts as a potential scatterer. The sum of all
dopants gives rise to a screened Coulomb potential  $V_i =V_0 f_i$ at site $i$ where
$f_i=\sum_s \exp(-r_{is}/\lambda)/r_{is}$, with $r_{is}$ the distance from
dopant $s$ to the lattice site $i$ in the plane, in units of
$\sqrt{2}a$, where $a$ is the Cu-Cu distance. The pair interaction $g_{ij}$
which enters the self-consistency condition for the
nearest neighbor $d$-wave order
parameter $\Delta_{ij}=g_{ij} \langle \hat{c}_{i\uparrow}
\hat{c}_{j\downarrow} - \hat{c}_{j\downarrow}
\hat{c}_{i\uparrow}\rangle$ is assumed constant,
i.e. $g_{ij}=g_0$ in model (i).
In model (ii) we assume that each oxygen dopant
locally modulates the pair interaction, which we model as $g_{ij}=g_0+\delta g
(f_i+f_j)/2$ with $f_i$ as defined above. In model (ii) the oxygen
dopants exert no conventional Coulomb potential, i.e., $V_i=0$.
Since potential scattering enters in the $\tau_3$-channel in Nambu
notation we will also call model (i) the $\tau_3$-scattering
model. Analogously we will term model (ii) the $\tau_1$-scattering
model since order parameter variations enter in the $\tau_1$-channel
in Nambu notation.

\section{Correlation between LDOS and dopants}

\begin{figure}[t]
\includegraphics[clip=true,width=.99\columnwidth]{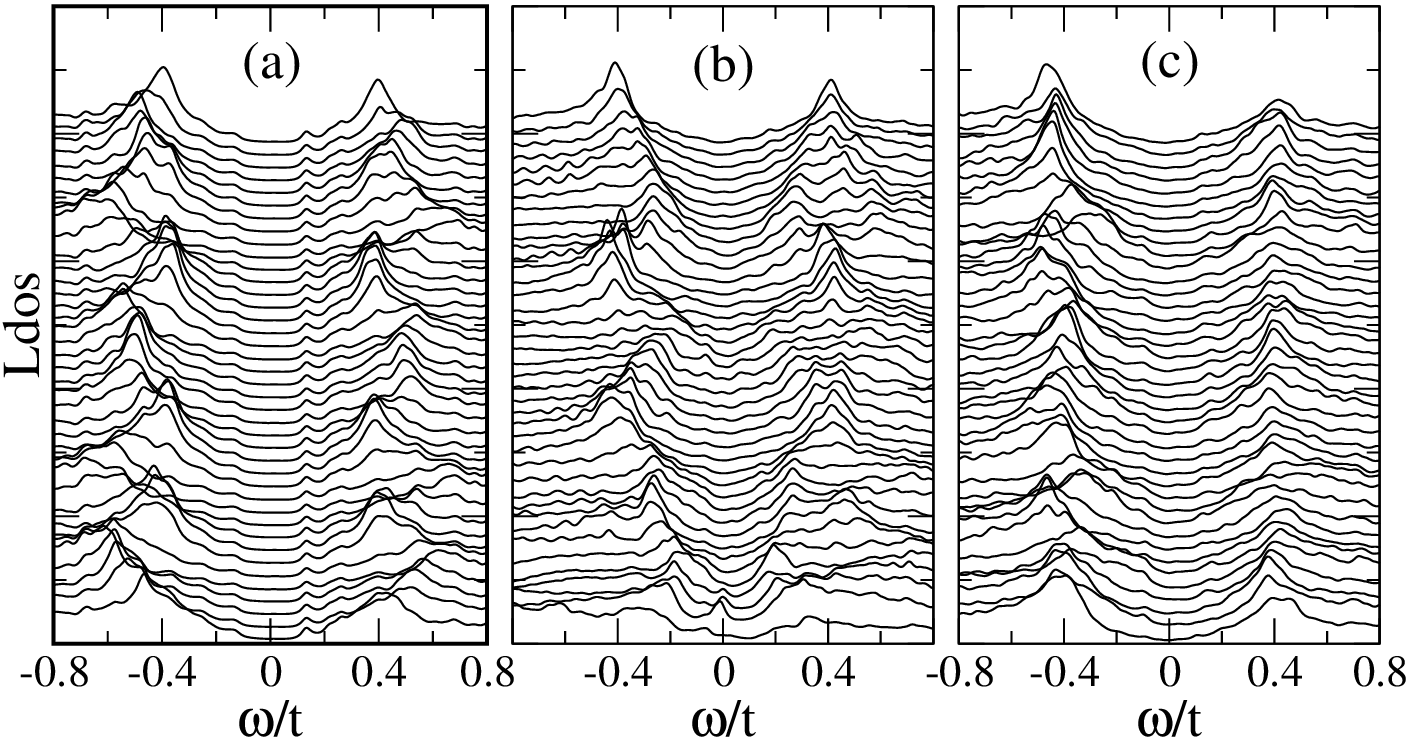}
\caption{LDOS along typical line cuts for (a) $\tau_1$-model with
$r_z=0.57$, $\lambda=0.5$,
  $\delta g=t$, (c) smooth $\tau_3$-model with $r_z=2$, $\lambda=2$,
  $V_0=1.5 t$ and (c) ``spiky'' $\tau_3$-model with $r_z=0.57$, $\lambda=0.5$,
  $V_0=t$.}
\label{fig:LDOSCuts}
\end{figure}

In Fig.~\ref{fig:LDOSCuts} the LDOS is shown for both the
$\tau_1$- and the $\tau_3$-scattering model along typical line
cuts. The LDOS has been obtained by solving self-consistently the
Bogoliubov-de Gennes equations resulting from
Eq.~(\ref{eq:hamiltonian}) on a $80 \times 80$ lattice rotated by
45 degrees with respect to the Cu-Cu bonds, i.e., for a system
containing $2\times 80 \times 80$ sites. Panel (a) shows the LDOS
for the $\tau_1$-scattering model. In good agreement with
experimental STS-data~\cite{Lang}, the LDOS along this linecut shows a strong
anticorrelation between height of coherence peaks and gap
magnitude. Furthermore, the LDOS is quite particle-hole symmetric
and fairly homogeneous at small energies, which is also in
favorable agreement with experimental findings. Panels (b) and (c)
show the LDOS for the $\tau_3$-scattering model in the limit of a
smooth potential variations, which arise if the dopants are poorly
screened, and in the limit of a ``spiky'' potential, which arises
when each dopant is well screened and exerts only a short ranged
potential. Obviously the $\tau_3$-model fails to reproduce the
experimental observations in both limits. In the smooth limit (b)
the height of the coherence peaks is positively correlated with
the gap magnitude in striking contrast to the experimental
observation~\cite{DavisScience05}. In the ``spiky'' limit (c) gap
modulations appear to be quite small, the LDOS lacks particle-hole
symmetry and the homogeneity at small energies is destroyed by the
formation of sub-gap resonances.

A more direct test for the proposed scattering models is to
consider the spatial correlation functions $C_{\rho g}$ ($C_{\rho V}$)
between the LDOS $\rho_i(\omega)$ and the modulated pair interaction $g_i$
(impurity potential $V_i$) which track the O dopant positions in our
model.  For example, the spatial 
correlation function between the LDOS $\rho_i(\omega)$ and the
impurity potential $V_i$ is defined as
\begin{equation}
C_{\rho V}(\omega)\!=\!\frac{\sum_i \left( \rho_i(\omega)
-\langle \rho (\omega) \rangle \right) \left( V_i-\langle
V \rangle \right)}{\sqrt{A_\rho A_V}}\,, \label{eq:LDOSOcorr}
\end{equation}
where $A_V = \sum_i \left( V_i - \langle
  V \rangle\right)^2$ and $\langle V \rangle$ is the average of $V$.

\begin{figure}[t]
\begin{minipage}{.49\columnwidth}
\includegraphics[clip=true,width=.99\columnwidth]{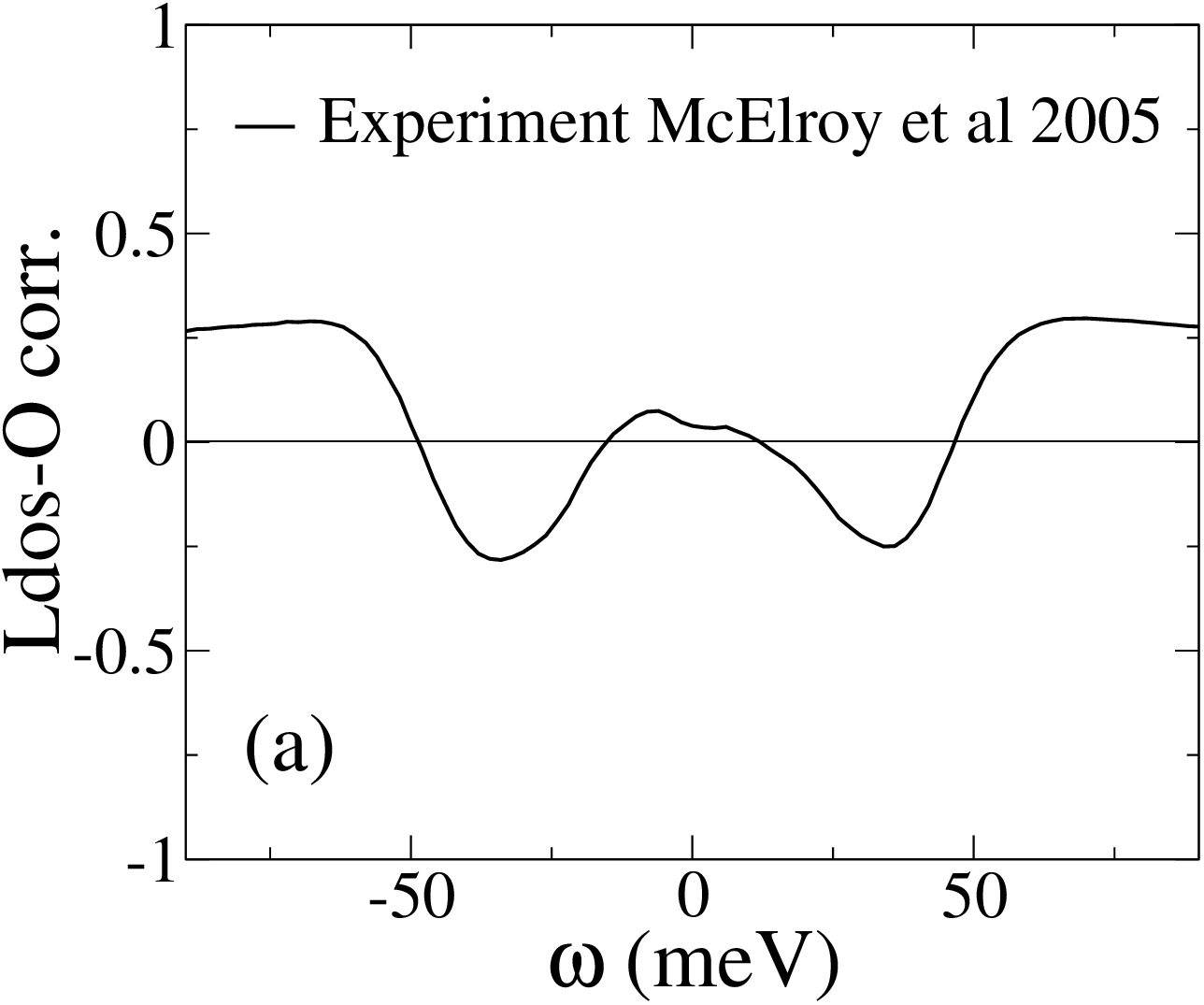}
\end{minipage}
\begin{minipage}{.49\columnwidth}
\includegraphics[clip=true,width=.99\columnwidth]{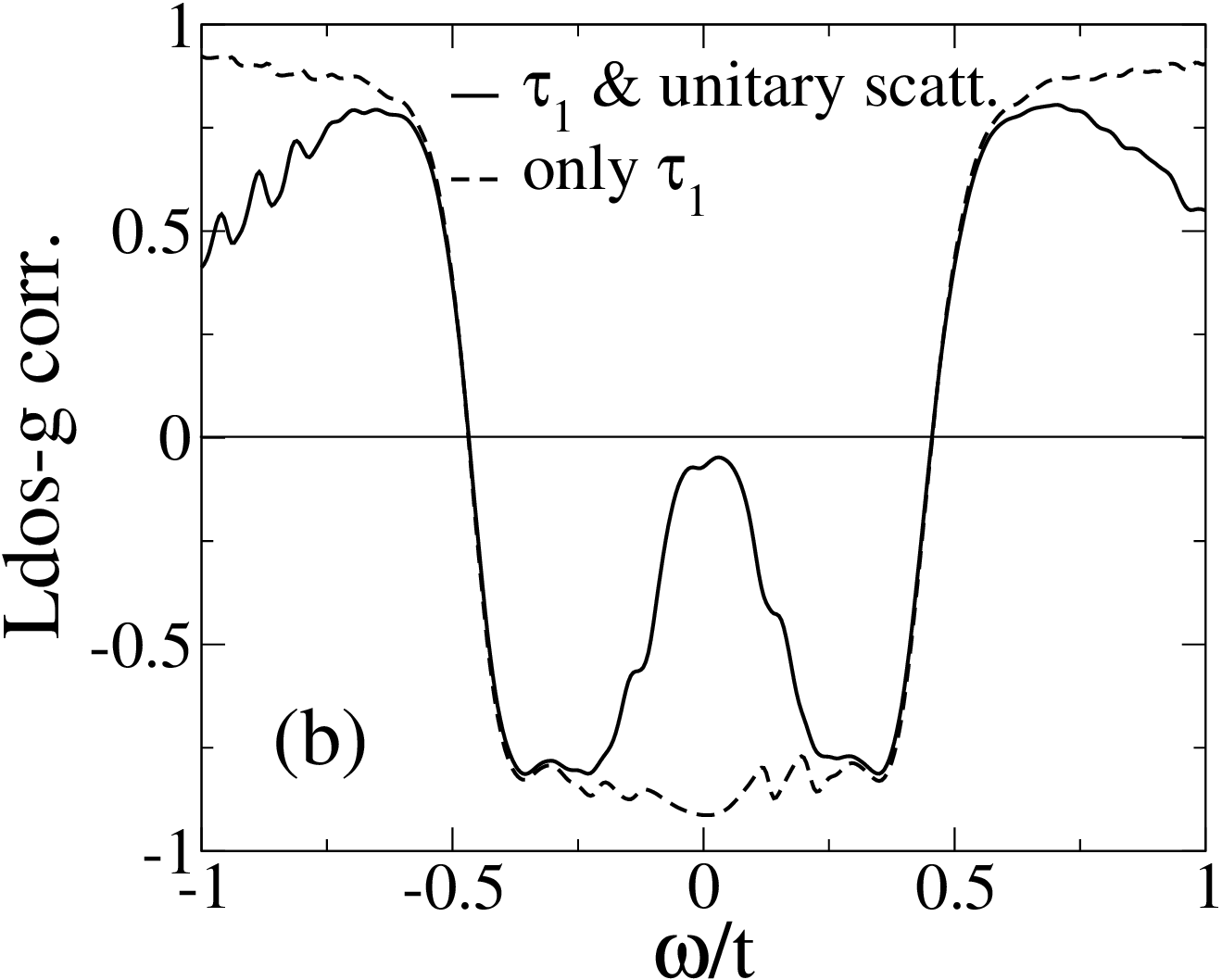}
\end{minipage}\\
\begin{minipage}{.49\columnwidth}
\includegraphics[clip=true,width=.99\columnwidth]{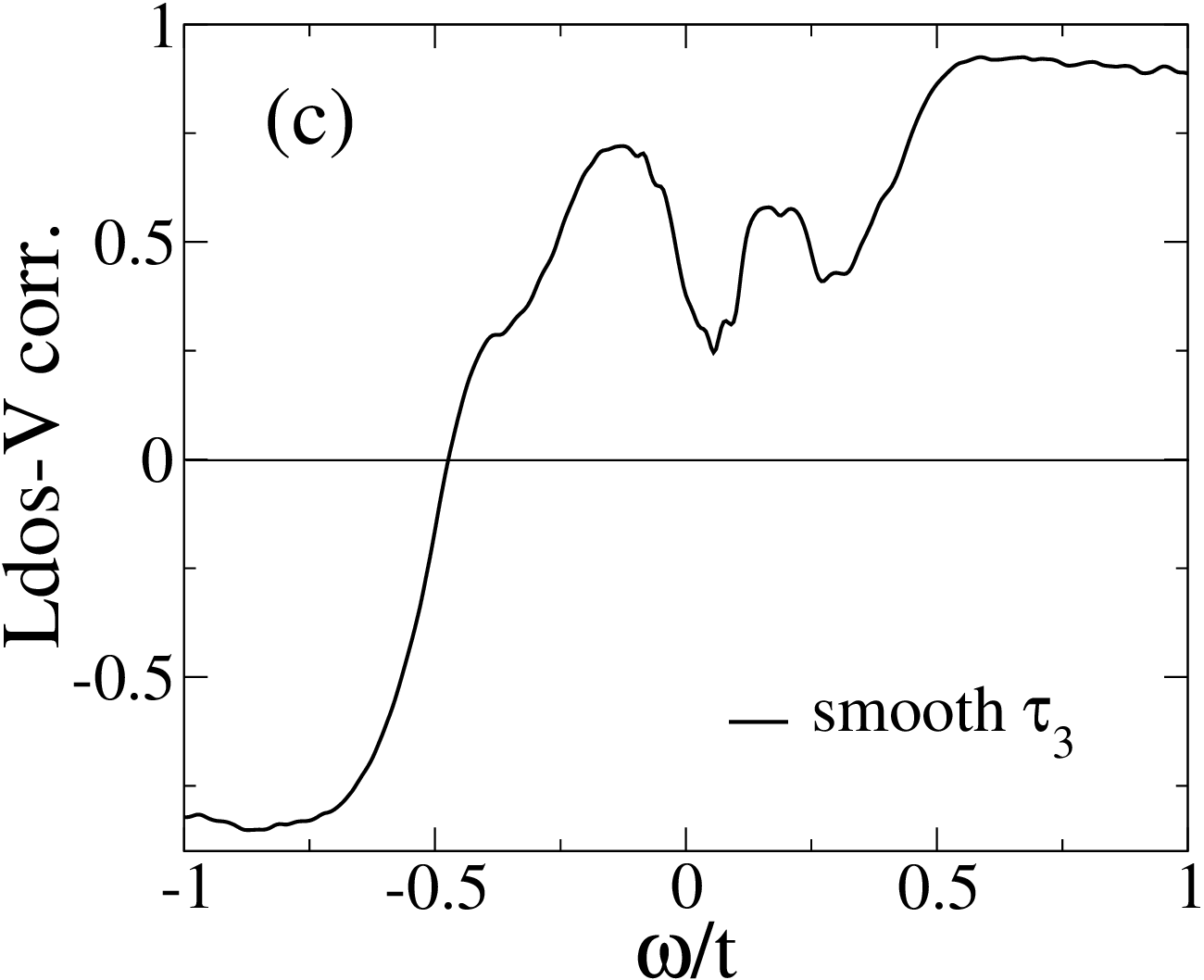}
\end{minipage}
\begin{minipage}{.49\columnwidth}
\includegraphics[clip=true,width=.99\columnwidth]{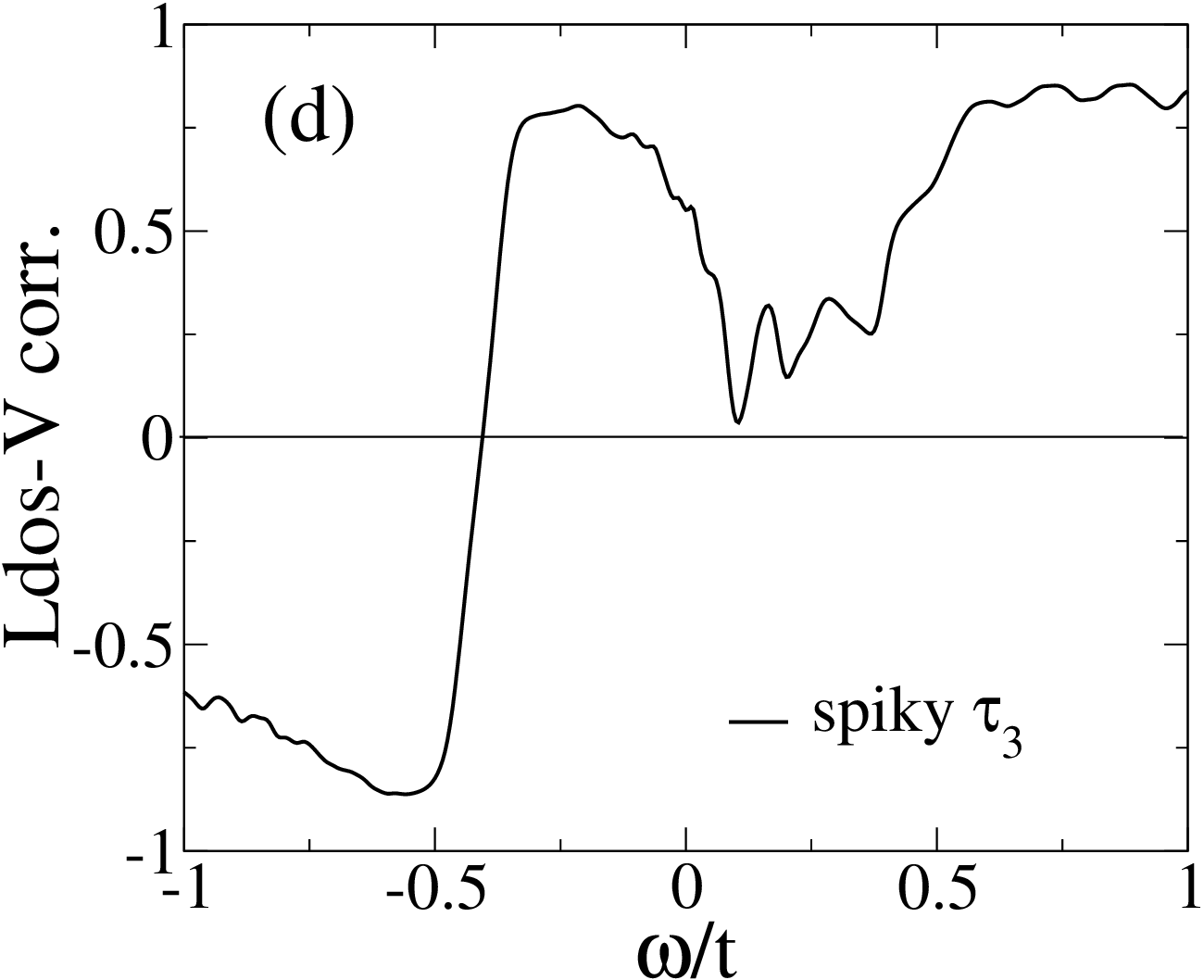}
\end{minipage}
\caption{Correlation function between LDOS and (a) position of the
oxygen dopant atoms (experimental) (b) pair interaction $g_i=g_0 +
\delta g f_i$ ($\tau_1$-model with $r_z=0.57$,
$\lambda=0.5$, $\delta g=t$), (c) impurity potential $V_i$
(smooth $\tau_3$-model with $r_z=2$, $\lambda=2$,
  $V_0=1.5 t$) and (d) impurity potential  $V_i$ (``spiky''
  $\tau_3$-model with $r_z=0.57$, $\lambda=0.5$, $V_0=t$).}
\label{fig:LDOSOcorr}
\end{figure}

The correlation functions for the $\tau_1$- and $\tau_3$-models
are shown in Fig.~\ref{fig:LDOSOcorr} in comparison with the
experimental result (panel (a)). The experimental correlation
function between the LDOS and the position of the oxygen dopant
atoms (a) is remarkably particle-hole symmetric with  maximal
negative values near $\pm$35 meV and maximal positive values near
$\pm$70 meV. This behavior of the correlation function reflects the
enhancement of the gap close to the oxygen dopant
atoms, i.e., the coherence peaks which correspond to additional
spectral weight move to higher energies. At small energies the
correlation almost disappears, which is a signature of the
homogeneity of the LDOS at small frequencies or of the presence of
other impurities uncorrelated with the oxygen dopants (see below).

The dashed line in Fig.~\ref{fig:LDOSOcorr}(b) shows the
correlation function $C_{\rho g}$ between the LDOS and the pair
interaction $g_i=g_{ii}=g_0 + \delta g f_i$ for the
$\tau_1$-scattering model. 
In good agreement with the experimental curve (a), the correlation
function is quite particle-hole symmetric, and is negative for
small energies and positive for large energies. 
In contradiction to experiment, however, the correlation function
remains negative and almost constant at small energies, in particular
it does not vanish for energies close to zero.
This seems to indicate that the Ldos tracks the local gap magnitude at
small energies as one would expect for a smoothly varying pair
potential, i.e., the slope of the LDOS at small energies decreases because 
the gap increases in regions of large pair interaction (close to the oxygen dopants).
%
We believe that this discrepancy with the
experimental data is due to the fact that the
real BSCCO-2212 system contains further sources of disorder
besides the oxygen dopant atoms. In particular, about 0.2\%
in-plane native defects are typically imaged on a BSCCO-2212
surface. Due to the resonances they generate near zero
bias~\cite{native} they are generally considered to be strong
pointlike scatterers. In order to mimic the real BSCCO-2212 system
these additional native defects have to be taken into
account.

The solid line in Fig.~\ref{fig:LDOSOcorr}(b) shows the
correlation function $C_{\rho g}$ between the LDOS and the pair
interaction for a system where 0.25\% strong
pointlike potential scatterers ($V=5t$) have been included in
addition to the dopants which only modulate the pair interaction. Obviously, the presence of a
small number of unitary scatterers  affects the LDOS primarily at
very small and at large energies. In particular it destroys the
correlation near zero bias. The qualitative agreement between the
solid line in panel (b) and the experimental curve in panel (a) is
almost perfect. Only the quantitative agreement is not as satisfactory
because  the correlations in our model calculation are by
a factor of 2-3 larger than experimental observations. We speculate
that this is due to additional disorder in the real BSCCO-2212
system which could e.g. stem from the random substitution of Bi at
the Sr sites~\cite{Eisaki}.

For comparison the correlation functions $C_{\rho V}$ between the LDOS and the
impurity potential $V_i$ for the $\tau_3$-scattering model
are shown in Fig.~\ref{fig:LDOSOcorr}(c) and (d) in the smooth and
``spiky'' limit respectively. Both cases are in striking
contrast to the experimental data displayed in panel (a). The
most drastic discrepancy is the asymmetry of the correlation
function in the $\tau_3$-model with respect to zero bias which
simply reflects the particle-hole asymmetry of conventional
potential scattering and is in clear contradiction to the
experimental observation. Only scattering at order-parameter
variations can account for the remarkable particle-hole symmetry
observed experimentally.

\section{Conclusions}

By comparing the correlation between the LDOS and the
oxygen dopants with experimental results we have provided
further evidence in favor of a model where the nanoscale
inhomogeneity in the local electronic structure of BSCCO-2212 is
caused by a dopant modulated pair interaction. Our results 
demonstrate also that the presence of native defects, which are imaged
as scattering resonances near zero bias on typical BSCCO-2212
surfaces, reduces the correlation between the LDOS and the dopant
atoms at small energies.

\vspace{.3cm}
{\it Acknowledgements}. 
Partial support for this research was provided by DOE
DE-FG02-05ER46236 (PJH), ONR N00014-04-0060 (PJH and BMA) and 
the Alexander von Humboldt Foundation (TSN).

\end{document}